\documentstyle[prl,aps,graphics,multicol] {revtex}

\begin{document}

% \draft

\title{Binding of Similarly Charged Plates: A Global Analysis}

\author{
André G. Moreira and Roland R. Netz}

\address{Max Planck Institute of Colloids
                and Interfaces, 14424 Potsdam, Germany}
\date{\today}
\maketitle

\begin{abstract}
Similarly and highly charged plates in the presence of 
multivalent counter ions attract each other, leading
to electrostatically bound states.
Using Monte-Carlo simulations we obtain the inter-plate
pressure in the global parameter space.
The equilibrium plate separation, where the pressure 
changes from attractive to repulsive, exhibits
a novel unbinding transition. 
A systematic and asymptotically exact
strong-coupling field-theory yields the bound state
from a competition between counter-ion entropy and electrostatic
attraction, in agreement with simple scaling arguments.
\end{abstract}


\begin{multicols}{2}

Experimentally, it has been known for a long time that highly
charged planar surfaces attract each other in the presence of
multivalent counter ions, inducing bound states.
This electrostatic binding restricts the swelling of calcium
clay particles\cite{Quirk} and leads to much reduced
water uptake of charged lamellar membrane systems\cite{Wenner}.
Attractive forces between charged surfaces have also 
been observed with the surface force apparatus\cite{Keki}.
Monte-Carlo simulations indeed confirmed that for a given
surface charge density there exists a threshold counter-ion valence 
above which attraction can be observed over some range
of plate separations\cite{Guldbrand}.

Theoretically, these observations came as a surprise, since 
the mean-field or Poisson-Boltzmann (PB) theory, which works
usually quite well for charged systems, predicts only 
repulsive forces between charged objects\cite{Andelman}. 
This contradiction
between observation and PB prediction resulted in an immense
theoretical activity, which aimed at understanding the simple model
of two uniformly and similarly charged planar 
surfaces interacting across a gap of width $d$ filled with 
point-like counter ions. 
Clearly, reality is much more complicated due to additional
interactions, but even this simple model, which 
we will consider in the following, is quite challenging.
A number of approaches were proposed which incorporate
counter-ion correlations that are neglected within PB.
The first were integral-equation 
theories\cite{Kjellander}, perturbative expansions around the 
PB theory\cite{Attard,Podgornik}, and local density-functional 
theory\cite{Stevens}, which compare well with 
simulation results and exhibit attraction.
If the two plates are far apart from each other, 
the counter-ion clouds can be viewed as condensed on the plates,
and the resulting simplified model can be solved within
a Gaussian\cite{Pincus} or harmonic-plasmon approximation\cite{Lau}.
These approaches either involve numerics and do not provide
much physical insight, or they are valid for large separations
and cannot be used to characterize the bound state. 

Of great significance is the fact that when the electrostatic
force is attractive, the equilibrium distance 
$d^*$ between the plates is smaller than the typical lateral distance 
$a$ between counter ions (as we will 
demonstrate later on), thus rendering a quasi-two-dimensional
layer of counter ions\cite{Rouzina}. 
In the first part of this paper, we  will use this fact and
present a scaling argument for the attraction between two plates,
valid for $d \ll a$.
 We will then demonstrate that
this scaling analysis is equivalent to the leading order
of a systematic field-theory, valid
in the strong-coupling (SC) limit (corresponding to
large plate-charge density
$\sigma$ or large counter-ion valence $q$), where it agrees with 
extensive Monte-Carlo (MC) simulations.
Our MC results span the complete parameter space.
Whenever attractive forces between the plates exist, they
induce a bound equilibrium state, which exhibits a novel
unbinding transition.

The simple scaling argument for attraction between charged plates
starts with partitioning the system into isolated counterions
sandwiched between two finite plate segments of area 
$A = q/2\sigma$. Neglecting ion-ion interactions should be valid
for $d \ll \sqrt{A}$.
Denoting the distance 
between the counterion and the plates as $x$ and $d-x$,
respectively, we obtain for the electrostatic interaction
energies, in units of $k_BT$ and for $d \ll \sqrt{A}$,  the results
$W_1=2 \pi \ell_B q \sigma x $ and
$W_2=2 \pi \ell_B q \sigma(d- x) $, respectively, 
as follows from the potential at an infinite charged wall
and omitting constant terms.
The sum of the two interactions is 
$W_{1+2}=W_1+W_2 = 2 \pi \ell_B q \sigma d $ which shows that
i) no pressure is acting on the counterion since the forces
exerted by the two plates exactly cancel and ii) that 
the counterion  mediates an effective attraction between the two
plates. The interaction between the two plates is 
proportional to the total charge on one plate, $A \sigma$, and
for $d \ll \sqrt{A}$ given by
$W_{12} = -2 \pi A \ell_B \sigma^2 d$.
Since the system 
is electroneutral, $q = 2 A \sigma$, the total energy is 
$W= W_{12}+W_1+W_2 = 2 \pi A \ell_B \sigma^2 d$, leading
to an electrostatic pressure $P_{el} = -\partial (W/A)/\partial d 
=-2\pi \ell_B \sigma^2$. The two plates attract each other.
The entropic pressure due to counter-ion confinement 
is $P_{en} = 1/ A d = 2\sigma/q d$.
The equilibrium plate separation is characterized by zero total 
pressure, $P_{tot}=P_{el}+P_{en}=0$, 
leading to an equilibrium plate separation
$d^* = 1/\pi \ell_B q \sigma$. 
By construction, the derivation for $d^*$ is valid only for $d^* < a$
(where $a$ is the average lateral distance between ions as defined
by $\pi a^2 = q/2\sigma$), equivalent to the condition $\Xi =
2 \pi \ell_B^2 \sigma q^3 >4$, i.e. for large values of
the coupling constant $\Xi$.
Surprisingly, these results
for $P_{tot}$ and $d^*$ become exact
in the SC limit $\Xi \rightarrow \infty$, 
as we will demonstrate in the
following. In fact, the attraction between charged plates,
as derived here for $d \ll a$, is conceptually simpler
than the PB result of repulsion, because the latter
case involves many-body effects.

To proceed with our systematic field theory, 
consider the partition function for $N$ counter ions
confined between two parallel plates at distance $d$ 
\begin{equation}
{\cal Z  }_N = \frac{1}{N!} \prod_{j=1}^N 
\int {\rm d} {\bf r}_j \theta(z_j) \theta(d-z_j)
{\rm e}^{- {\cal H}}
\end{equation}
where the Heavyside function is defined by $\theta(z)=1$
for $z>0$ and zero otherwise. Introducing the counter-ion
density operator
$\hat{\rho}({\bf r}) = \sum_{j=1}^N \delta({\bf r}-{\bf r}_j)$
the Hamiltonian can be written as
\end{multicols}
\begin{equation}
{\cal H} = \frac{\ell_B}{2} \int {\rm d}{\bf r}{\rm d}{\bf r}'
[q \hat{\rho}({\bf r})-\sigma \delta(z) -\sigma \delta(d-z)]
v({\bf r}-{\bf r}') 
[q \hat{\rho}({\bf r}')-\sigma \delta(z') -\sigma \delta(d-z')]
-\int {\rm d}{\bf r}\;  \hat{\rho}({\bf r}) h ({\bf r})
\end{equation}
where $v({\bf r}) =1/r$ is the Coulomb interaction
and the field $h$ has been added to calculate density distributions
later on. The characteristic length scales
are the Bjerrum length $\ell_B= e^2/4\pi \varepsilon k_BT$ and
the Gouy-Chapman length $\mu = 1/ 2\pi \ell_B q \sigma$,
which measure the distance at which the interaction between
two unit charges and between
a counter ion and a charged wall reach thermal energy, respectively.
Rescaling all lengths by the 
Gouy-Chapman length according to ${\bf r} = \mu \tilde{\bf r}$
and $d= \mu \tilde{d}$, the Hamiltonian becomes 
\begin{equation}
{\cal H} = \frac{1}{8 \pi^2 \Xi} 
\int {\rm d}\tilde{\bf r}{\rm d}\tilde{\bf r}'
[2\pi\Xi \hat{\rho}(\tilde{\bf r})- \delta(\tilde{z})
-\delta(\tilde{d}-\tilde{z})]
v(\tilde{\bf r}-\tilde{\bf r}') 
[2\pi\Xi \hat{\rho}(\tilde{\bf r}')-\delta(\tilde{z}') 
-\delta(\tilde{d}-\tilde{z}')]
-\int {\rm d}\tilde{\bf r}\;  \hat{\rho}(\tilde{\bf r}) 
h (\tilde{\bf r})
\end{equation}
and thus only depends on the coupling parameter
$\Xi = 2 \pi q^3 \ell_B^2 \sigma$.
At this point we employ a Hubbard-Stratonovitch transformation,
 similar to previous implementations of a field theory for charged 
systems\cite{Netz}, followed by a Legendre transformation  to the
grand-canonical ensemble,
${\cal Q} = \sum_N \lambda^N {\cal Z}_N$, introducing the fugacity
$\lambda$.
The inverse Coulomb operator follows from
Poisson's law as
$v^{-1}({\bf r}) = -\nabla^2 \delta({\bf r})/4 \pi$,  which leads to 
\begin{equation} \label{Q}
{\cal Q  } =  \int \frac{{\cal D}\phi }{{\cal Z}_v} 
\exp \left\{ - \frac{1}{8\pi \Xi} \int {\rm d}\tilde{\bf r} \left[
[\nabla \phi(\tilde{\bf r})]^2
-4 \imath \delta(\tilde{z}) \phi(\tilde{\bf r})
-4 \imath \delta(\tilde{d}-\tilde{z}) \phi(\tilde{\bf r})
-4\Lambda  \theta(\tilde{z}) \theta(\tilde{d}-\tilde{z}) 
{\rm e}^{h(\tilde{\bf r}) -
\imath \phi(\tilde{\bf r})} \right] \right\}
\end{equation}
\begin{multicols}{2}
\noindent
where we introduced the notation ${\cal Z}_v = \sqrt{\det v}$
and the rescaled fugacity $\Lambda$ is defined by
$\Lambda = 2 \pi \lambda \mu^3 \Xi = \lambda/(2 \pi \sigma^2 \ell_B)$.
The expectation value of the
counter-ion density, $ \rho(\tilde{\bf r}) $, follows 
 by taking a functional derivative with respect
to the generating field $h$, $ \rho(\tilde{\bf r}) =
\delta \ln {\cal Q} / \delta h(\tilde{\bf r})  \mu^3$,
giving rise to
\begin{equation} \label{dens}
\tilde{\rho}(\tilde{\bf r}) =
\frac{ \rho(\tilde{\bf r})  }{2 \pi \ell_B \sigma^2}=
 \Lambda \langle {\rm e}^{-\imath
\phi(\tilde{z})} \rangle.
\end{equation}
The normalization condition for the counter-ion distribution,
$\mu \int {\rm d} \tilde{z} \rho(\tilde{z}) = 2 \sigma/ q$,
which follows directly from the definition of the grand-canonical 
partition function, leads to
\begin{equation} \label{norm}
\Lambda \int_0^{\tilde{d}} {\rm d} \tilde{z} \langle {\rm e}^{-\imath
\phi(\tilde{z})} \rangle = 2.
\end{equation}
This is an important equation since it shows that the
expectation value of the
fugacity term in Eq.(\ref{Q}) is bounded and of the order
of unity.
Let us first repeat the saddle-point analysis, which, because
of the structure of the action in Eq.(\ref{Q}), should be valid for
$\Xi \ll 1$. The saddle-point equation reads
$  \partial^2 \phi(\tilde{z})/\partial  \tilde{z}^2 =
2 \imath \Lambda {\rm e}^{ - \imath \phi(\tilde{z})}$. 
The solution of this differential equation is
$ \imath \phi(\tilde{z}) = 2 \ln 
\cos \left(\Lambda^{1/2} [\tilde{z}- \tilde{d}/2] \right)$.
The normalization condition leads to the equation
$\Lambda^{1/2} \tan[\tilde{d} \Lambda^{1/2}/2] =1$,
which is solved by $\Lambda \simeq 2/\tilde{d}$ for
$\tilde{d} \ll 1$  and $\Lambda \simeq \pi^2/4\tilde{d}^2$
for $\tilde{d} \gg 1$. From Eq.(\ref{dens}),
the rescaled density distribution of counter ions is given by the 
well-known PB result
\begin{equation} \label{PB}
\tilde{\rho}(\tilde{z}) = 
1/\cos^2 \left(\Lambda^{1/2} [\tilde{z}- \tilde{d}/2] \right).
\end{equation}
Let us now consider the opposite limit, when the 
coupling constant $\Xi$ is large. In this case, the 
saddle-point approximation breaks down, since the prefactor in front
of the action in Eq.(\ref{Q}) becomes small. 
Since the fugacity term is bounded, as evidenced by Eq.(\ref{norm}), one can
expand the partition function (and also all expectation values)
in powers of $\Lambda/\Xi$ (which is equivalent to a virial expansion).
For the expectation value determining the density Eq.(\ref{dens}) 
the leading two orders in the virial expansion are
\[
\langle {\rm e}^{-\imath
\phi(\tilde{z})} \rangle =
 {\rm e}^{-\Xi v(0)/2} 
 +\frac{\Lambda {\rm e}^{-\Xi v(0)}} {2 \pi \Xi}
  \int {\rm d}{\bf r}\left( {\rm e}^{-\Xi 
 v({\bf r} - \tilde{\bf r})} -1\right) .
\]
The normalization condition Eq.(\ref{norm}) 
can then be solved by an expansion of the fugacity in inverse powers
of the coupling constant, $\Lambda = \Lambda_0 + \Lambda_1 /\Xi + \ldots$.
We obtain $ \Lambda_0=  2 {\rm e}^{\Xi v(0)/2}/\tilde{d}$
and thus the density distribution is 
(in agreement with our scaling analysis)
to leading order indeed a constant
given by

\begin{figure}
%\vspace*{3mm}
\begin{center}
\resizebox{8.7 cm}{!}{
\includegraphics{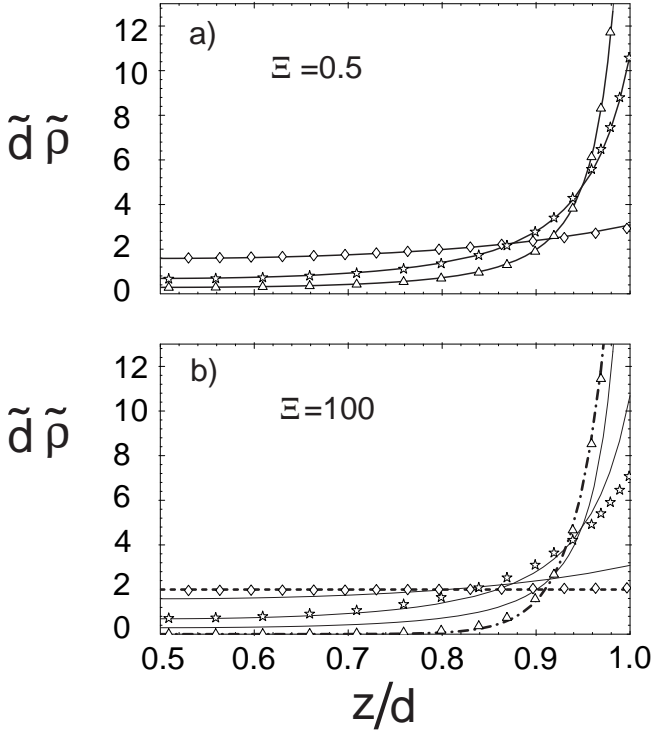}
}
\end{center}
\caption{
\narrowtext
MC results for the rescaled counter-ion density
$\tilde{d} \tilde{\rho}$ as a function
of the rescaled distance from the wall $z/d $
in the a) PB limit for $\Xi=0.5$ and in the b)
SC limit for $\Xi=100$ for various plate separations
$\tilde{d}=d/\mu = 1.5 $ (open diamonds),
$\tilde{d}=10$ (open stars), and $\tilde{d}=30$ (open triangles).
In a) MC  results agree well with the corresponding
PB predictions  (Eq.(\ref{PB}), solid lines), whereas in b) 
results for $\tilde{d} =1.5 $ agree
with the asymptotic SC prediction, Eq.(\ref{strong}) (dashed line)
and for $\tilde{d}=30$
with a double-exponential curve (see text).}
\end{figure}

\begin{equation} \label{strong}
\tilde{\rho}(\tilde{z}) = 2/\tilde{d} +{\cal O}(\Xi^{-1})
\end{equation}
In Fig.1a we show counter-ion density profiles obtained using
MC simulations \cite{Lekner}
 for small coupling parameter $\Xi=0.5$
for various plate distances,
which are well described by the PB profiles  Eq.(\ref{PB}) 
shown as solid lines. Fig.1b shows that for
$\Xi=100$ PB (thin solid lines) is inadequate\cite{com}. 
As suggested by our scaling analysis,
the asymptotic SC result Eq.(\ref{strong}) should be valid 
for  $d/a=\tilde{d}/\Xi^{1/2}<1$ only, since otherwise ion-ion 
interactions become important.
For $\tilde{d}=3/2$ (open diamonds) we find $d/a = 0.15$, 
and indeed Eq.(\ref{strong}) is accurate. For 
$\tilde{d}=10$ (open stars) we find $d/a = 1$, the density profile
is neither described by Eq.(\ref{strong}) nor (\ref{PB}).
Finally,  for $\tilde{d}=30$ (open stars) we find $d/a = 3$,
the two layers are decoupled and the density profile is described
by a double exponential $\tilde{\rho}(\tilde{z}) = 
({\rm e}^{-\tilde{z}}+{\rm e}^{\tilde{z}-\tilde{d}})/
(1-{\rm e}^{-\tilde{d}})$ (dashed-dotted line),
which is the superposition
of the density profiles of two isolated charged surfaces
in the SC limit\cite{Moreira}.
The crossover from PB to SC is demonstrated in Fig.2,
where we plot density profiles for fixed plate
separation $\tilde{d}=2$ for various coupling parameters
$\Xi$.

\begin{figure}
%\vspace*{3mm}
\begin{center}
\resizebox{8.7 cm}{!}{
\includegraphics{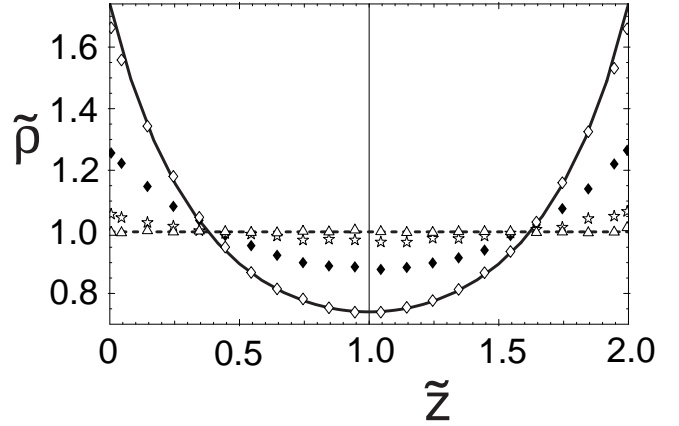}
}
\end{center}
\caption{
\narrowtext
MC results for rescaled counter-ion density profiles
$\tilde{\rho}=\rho/2 \pi \ell_B \sigma^2$ 
for fixed plate separation $\tilde{d} = d/\mu =2 $ 
as a function of the rescaled distance $\tilde{z} = z/\mu$
from one wall. Symbols correspond to
coupling parameters $\Xi=0.5$ (open diamonds),
$\Xi=10$ (filled diamonds), $\Xi=100$ (open stars), 
and $\Xi=10^5$ (open triangles), exhibiting clearly the crossover
from the PB prediction (solid line, Eq.(\ref{PB}))
to the SC prediction (broken line, Eq.(\ref{strong})).}
\end{figure}

Using the contact value theorem, the pressure 
$P$ between the two plates, which follows from the
partition function via $P=\partial \ln {\cal Q}_\lambda/
A \mu \partial \tilde{d}$, is related to the counter ion 
density at a plate by\cite{Guldbrand,Netz}
\begin{equation} \label{contact}
\tilde{P}=\frac{P}{2 \pi \ell_B \sigma^2}=
\tilde{\rho}(\tilde{d})-1.
\end{equation}
The first term is the entropic pressure due to counter-ion
confinement, the second term is due to electrostatic interactions
between the counterions and the charged plates.
Numerically, the contact ion density  $\tilde{\rho}(\tilde{d})$
is obtained from the density
profiles by extrapolation. In Fig.3 we show numerical pressure
data for various values of $\Xi$. Attraction (negative
pressure) is obtained for $\Xi>10$. The numerical pressure for 
$\Xi=0.5$ (open diamonds) agrees well with the PB prediction (solid line),
which from Eqs.(\ref{contact}) and (\ref{PB}) is given by
$\tilde{P} = \Lambda$ with $\Lambda$ determined by
$\Lambda^{1/2} \tan[\tilde{d} \Lambda^{1/2}/2] =1$. The small distance
range of most data, and the complete pressure data for
$\Xi = 10^5$ (open triangles) are well described by the SC prediction
(broken line). It results from
combining Eqs.(\ref{contact}) and (\ref{strong}) and is given by
$\tilde{P} = 2/\tilde{d}-1$, from which  
the equilibrium separation, which corresponds to the minimum of
the effective plate-plate interaction, is obtained as
$\tilde{d}^* = 2$. Incidentally, this is exactly the 
scaling prediction for the pressure derived in the beginning of this paper.

\begin{figure}
%\vspace*{3mm}
\begin{center}
\resizebox{8.7 cm}{!}{
\includegraphics{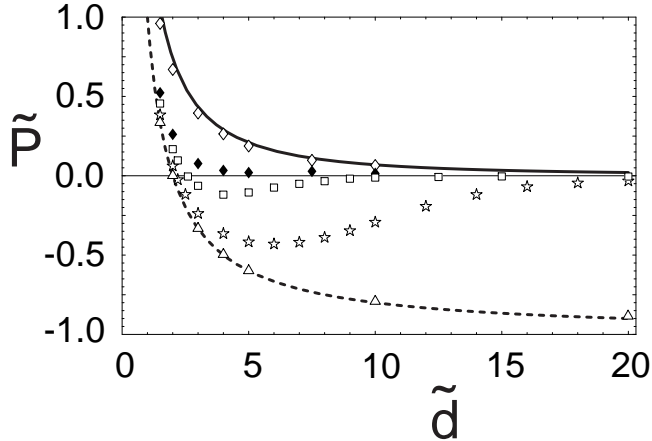}
}
\end{center}
\vspace*{-2mm}
\caption{
\narrowtext
MC results for the rescaled pressure 
$\tilde{P} = P/2 \pi \ell_B \sigma^2$ 
as a function of the rescaled plate separation
$\tilde{d} = d/\mu$
for the same parameter values
as in Fig.2 (and $\Xi=20$, open squares), 
compared with the PB prediction $\tilde{P}=\Lambda$
(solid line)
and the SC prediction $\tilde{P} =2/\tilde{d}-1$ (broken line).}
\end{figure}
\vspace*{-2mm}

Finally, combining all pressure data, we obtain the 
global phase diagram shown
in Fig.4, featuring two regions where the inter-plate pressure
is attractive and repulsive.
The dividing line between those regions,
which corresponds to the equilibrium plate separation 
$\tilde{d}^*$ 
in the bound state, is determined over 4 decades 
of the coupling constant $\Xi$. 
In the limit of large coupling constants, the phase boundary
saturates at $\tilde{d}^*=2$, in agreement with our scaling
argument and the leading result of our SC theory.
The threshold coupling constant to observe attraction between
charged plates is $\Xi^* \approx 10$. As $\Xi \rightarrow \Xi^*$
from above, the equilibrium plate separation 
diverges continuously to infinity. This constitutes a novel
unbinding transition, which experimentally could be observed
with charged lamellar or clay systems 
by raising the temperature.

The plate separation equals the 
lateral ion-distance on a line in Fig.4 determined by
$d/a = \tilde{d}/\Xi^{1/2} =1$,
which crosses the equilibrium-separation line at 
$\tilde{d} \approx 3$. 
For the most part of the equilibrium-separation line in Fig.4 
the counter-ion distribution is therefore indeed two-dimensional, 
and many-ion effects can be neglected except very close
to  $\Xi^*$. 
Correlations between counter ions, except the
lateral exclusion correlation  which keeps ions apart,
are therefore mostly unimportant in the bound state.
This explains
why the simple single-ion scaling argument 
advanced in the beginning, which turns out to be exact in the limit
$\Xi \rightarrow \infty$, works so well.
More detailed comparison between MC and
next-leading terms of our SC theory, which are
equivalent to higher-order virial terms,
together with a detailed discussion of the significance
of Wigner crystallization (which occurs at $\Xi \simeq 15600$
in the limit $\tilde{d} \rightarrow 0$) will 
be published shortly.

\begin{figure}
%\vspace*{3mm}
\begin{center}
\resizebox{8.7 cm}{!}{
\includegraphics{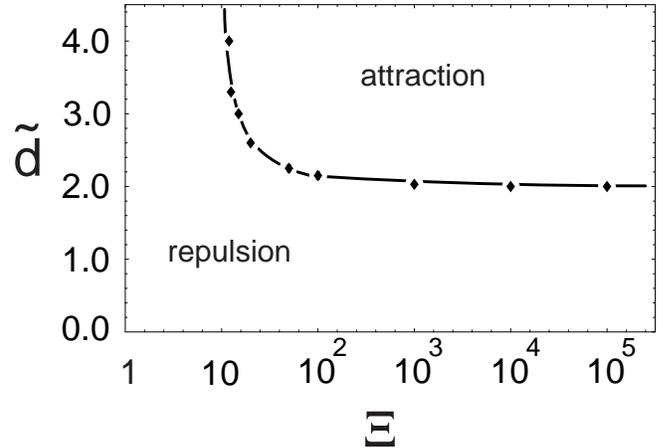}
}
\end{center}
\vspace*{-2mm}
\caption{
\narrowtext
Global phase diagram featuring the regions of repulsive
and attractive pressure as a function of the rescaled plate separation 
$\tilde{d}=d/\mu$ and coupling strength $\Xi$. The dividing line denotes
the equilibrium plate separation $\tilde{d}^*$, which 
saturates at $\tilde{d}^*=2$ for $\Xi \rightarrow \infty$
and which diverges as $\Xi$ approaches
the critical value $\Xi^* \approx 10$ from above.}
\end{figure}

\end{multicols}


\begin{thebibliography}{10}
\bibitem{com}
Experimentally, a coupling parameter $\Xi= 100$, which is 
quite close to the SC limit (see Fig.4),
is reached with divalent ions for a surface charge density
$\sigma \approx 3.6 nm^{-2}$, which is feasible with compressed
charged monolayers, and with trivalent counter ions for 
$\sigma \approx 1 nm^{-2}$, which is a typical value.
\end{thebibliography}
\end{document}